# The Wide-Field Imaging Interferometry Testbed[1]

Xiaolei Zhang,
Raytheon ITSS / NASA Goddard Space Flight Center
Lee Feinberg, David Leisawitz, Douglas B. Leviton, Anthony J. Martino, and John C. Mather
NASA Goddard Space Flight Center
Greenbelt, MD  20771
301-477-3562
zhang@stars.gsfc.nasa.gov

*Abstract*—We are developing a Wide-Field Imaging Interferometry Testbed (WIIT) in support of design studies for NASA's future space interferometry missions, in particular the SPIRIT and SPECS far-infrared/submillimeter interferometers.  WIIT operates at optical wavelengths and uses Michelson beam combination to achieve both wide-field imaging and high-resolution spectroscopy. It will be used chiefly to test the feasibility of using a large-format detector array at the image plane of the sky to obtain wide-field interferometry images through mosaicing techniques.  In this setup each detector pixel records interferograms corresponding to averaging a particular pointing range on the sky as the optical path length is scanned and as the baseline separation and orientation is varied.  The final image is constructed through spatial and spectral Fourier transforms of the recorded interferograms for each pixel, followed by a mosaic/joint-deconvolution procedure of all the pixels.  In this manner the image within the pointing range of each detector pixel is further resolved to an angular resolution corresponding to the maximum baseline separation for fringe measurements.

We present the motivation for building the testbed, show the optical, mechanical, control and data system design, and describe the image processing requirements and algorithms.  WIIT is presently under construction at NASA's Goddard Space Flight Center.

TABLE OF CONTENTS

## 1. INTRODUCTION

The proposed Submillimeter Probe of the Evolution of Cosmic Structure (SPECS, see Mather et al. 2000) and SPace InfraRed Interferometric Telescope (SPIRIT) missions are intended to observe the predominant far-infrared and submillimeter emissions in the universe, and to study how the first galaxies and structures formed and evolved.  Ideally, we would like to obtain images in the far-infrared covering the same spatial extent (~ several arcminutes) and with the same angular resolution (~ tens of milliarcseconds) as the Hubble Deep Fields.  This corresponds to roughly 10,000 x 10,000 resolution elements per image.  Furthermore, to support high spatial resolution interferometry and to study line emissions in galaxies, we will need spectral resolution capability.

The coming of age of large-format, background-limited far-infrared detector arrays calls for an exploration of the interferometer configurations that are best coupled to the characteristics of the detectors for acquiring wide-field, high spatial- and spectral-resolution images.  The SPECS and SPIRIT instruments, as currently envisioned, are both Michelson spatial-spectral interferometers with 2 or 3 collector mirrors surrounding a central hub, and are capable of varying baselines so as to achieve full spatial frequency coverage.  In order to establish the feasibility of using this configuration to achieve high-fidelity wide field imaging, a laboratory testbed is proposed for construction at NASA's Goddard Space Flight Center (Leisawitz et al. 1999).  This testbed will be a scaled-down version of the future SPIRIT and SPECS in size, operating at optical wavelengths, and using the same kind of detection schemes as will be used for the space interferometers.  In what follows, we describe the principles, the design and the objectives of the testbed, and report the progress to date on its construction and on understanding its expected performance.

## 2. WIDE FIELD IMAGING IN MICHELSON BEAM COMBINATION MODE

In the practice of radio interferometry, the correlation and interference of signals from the different antennas are done post-detection. In optical interferometry, however, the interference of signals from the different optical paths are done naturally by overlapping the beams in space, followed by the detection of the resulting fringes either in space or in time. Of the two widely used image-plane optical beam combination schemes, i.e. the Fizeau and the Michelson, the conventional view is that the Fizeau should be used for wide field applications, since the off-axis performance of a Fizeau-type interferometer is limited only by the aberrations in the optics. The so-called Michelson stellar interferometer, on the other hand, has been shown to have a much more limited field-of-view, due to the fact that its off-axis point-spread-function is in general different from the on-axis one, resulting from the different rates with which the fringe-envelope and the fringes themselves move as the off-axis angle increases (Traub 1986; Labeyrie 1996).

---

[1] Standard IEEE copyright which, for 2001, will be "O-7803-6599-2/01/$10.00 @ 2001 IEEE"

The field-of-view limitation of the Michelson stellar interferometer, however, does not afflict the so-called traditional Michelson spatial interferometer which implements a pupil-plane beam combination scheme. In this scheme, a beam splitter is used to combine pupil images at zero spatial separation. Furthermore, an imaging lens following the beam combiner can be used to map the rays from different sky directions into the different spatial locations on the focal plane. In this scheme the fringes corresponding to each sky direction are actually formed in time, or in the delay direction as the path length between the two interferometer arms changes. In this way the different rates for fringe-movement and fringe-envelope-movement no longer cause problems, since the fringe-movement is now recorded in the temporal direction for each sky direction. It is such a pupil-plane beam combination, image plane detection scheme that we plan to adopt for WIIT (Figure 1), and eventually also for SPECS and SPIRIT.

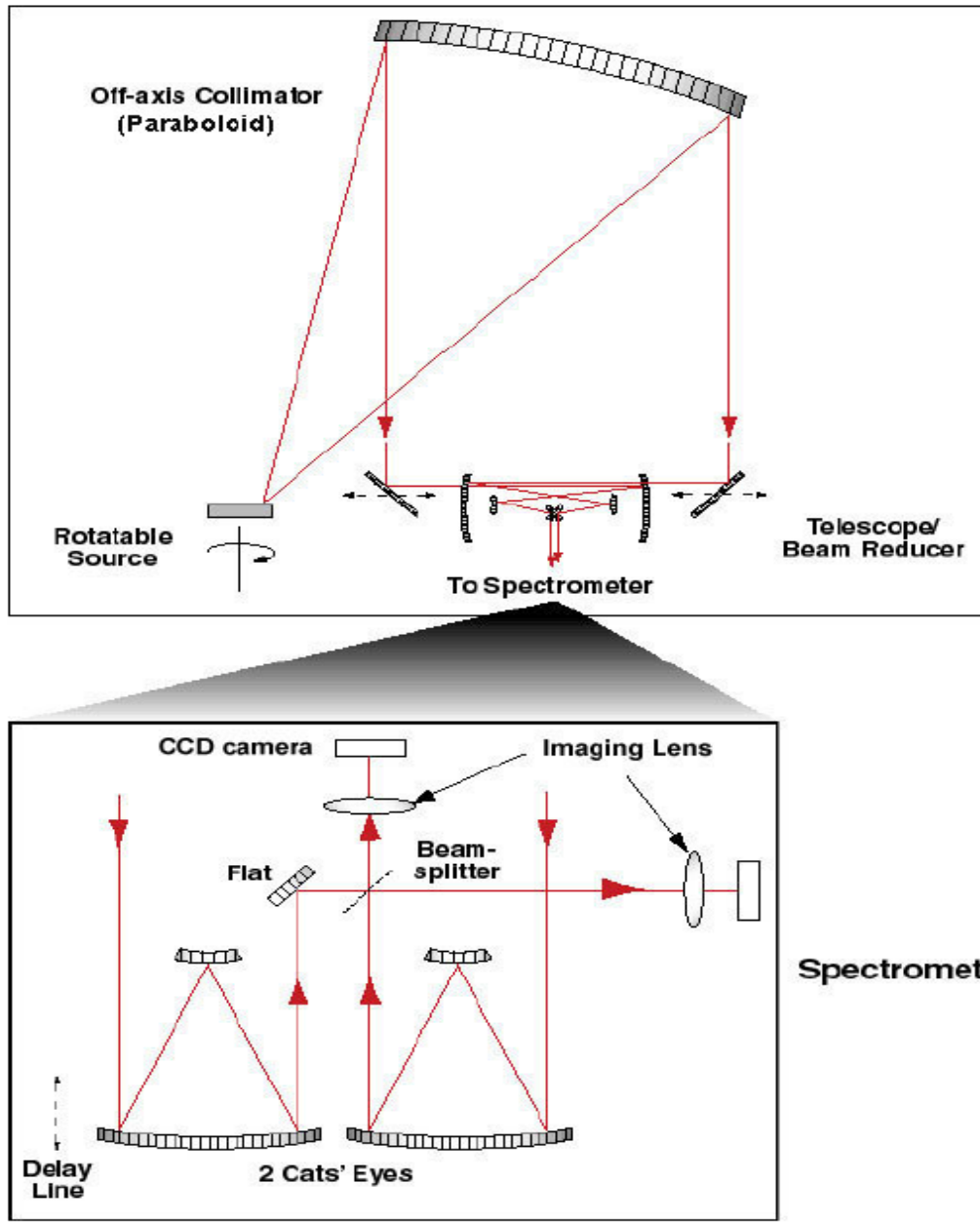

Figure 1: WIIT Optics

## 3. CONCEPTUAL DESIGN AND OBJECTIVES OF WIIT

In Figure 1, the schematics of the WIIT optics design is shown. A parabolic collimator produces parallel beams from a source located at its focus, simulating the "infinity" condition corresponding to a source on the sky. Two flat mirrors collect two parcels of the collimated wavefront, and after beam size reduction by the beam-reducer optical train and path length compensation by the cat's eye delay line, these pupil plane beams are combined through a beam splitter to produce two output beams. Each is further imaged by a dielectric lens before final detection at the focal plane of the imaging lens. The spatial frequency (the so-called u-v) plane is sampled through the combination of the variation of the baseline length by the two linear translation stages carrying the collector mirrors, and the variation of the effective baseline orientation by the rotary stage carrying the source.

We plan to place a CCD array detector at the focal plane of the lens, with the focal length of the lens properly chosen to match the spacings of the CCD pixels such that the pixels Nyquist sample the primary beam of the collector mirrors. As the more rigorous analysis later will show, this arrangement allows the interferometer images acquired by individual detector pixel to be "mosaiced" together. It also allows the recovery of a portion of the low spatial frequency flux that is not directly measured by a single pixel interferometer. The final linear resolution corresponds to the maximum baseline separation, and the linear resolution element per image is given by the number of linear CCD pixels times the number of interferometry resolution elements per CCD pixel. For the actual WIIT design (Zhang 2000a), this number is comparable to what is intended for SPECS, i.e. 10,000 linear resolution elements.

The main objectives of the testbed are the following:

1. To establish the feasibility of the basic design configuration. The actual demonstration of wide-field imaging using a Michelson interferometer has never been previously attempted, so the validity of this approach needs to be shown.
2. To study the optical performance of the design. The optics needs to be either aberration free both on- and off-axis over a wide field of view, or else an effective aberration compensation scheme needs to be devised in the optical setup or in the data reduction procedure.
3. To study the many technical issues of data acquisition such as metrology, absolute phase reference, long stroke delay and on-the-fly data taking. Effective calibration of the data also needs to be investigated.
4. To explore the various data analysis algorithms, especially the mosaic image-reconstruction and joint-deconvolution procedures developed by the radio astronomy community over the past decades.
5. To study the effective management of the large amount of data obtained by the wide-field, high resolution Michelson spatial and spectral interferometer, and to compare the efficiency of various data analysis algorithms.

During actual measurement, we will record fringes measured by each detector pixel as we vary the delay-line length. The white-light fringe peak for each sky direction will occur at different delay-line settings, the actual offset depending also on the baseline used for the particular measurement. The fringes are recorded at least twice per cycle for the shortest wavelength of interest. The absolute phase of the acquired fringes is referenced to a point source on axis. A spatial and spectral Fourier transform followed by a mosaic/joint-deconvolution procedure produces high-resolution wide-field images for each spectral channel.

## 4. MATHEMATICAL FORMULATION

Apart from the order of beam combination and signal detection, the method we use for interferometric data acquisition is very similar to that used in radio astronomy. In particular, both the optical and radio versions of the Michelson interferometers are designed to measure the so-called complex visibility function V, or the correlation between two complex signals $u_1$ and $u_2$. In radio, the correlation function $<u_1.u_2>$ is measured by a complex correlator. In optics, the detector measures effectively $<(u_1+u_2)^2> = <u_1^2+u_2^2+2u_1.u_2>$, and after subtracting the DC component $<u_1^2+u_2^2>$, the correlation function $<u_1.u_2>$ again results. Therefore, many useful mathematical tools developed in the practice of radio interferometry can be adapted to the optical case.

For a source confined to a small region of the sky surrounding coordinate center (0,0), the source brightness distribution B(l,m) near (0,0) and the visibility function V(u,v) measured by a Michelson interferometer with pointing and phase reference centers both at (0,0) can be shown to satisfy a two-dimensional Fourier transform relation

$$A(l,m).B(l,m) = \int_{-\infty}^{\infty}\int_{-\infty}^{\infty} V(u,v)e^{2\pi i(ul+vm)} du dv \quad (1)$$

where A(l,m) is the primary beam response of the collector mirror, and, in our case, it also incorporates the detector pixel function which produces an effective illumination on the collector mirror. In the above expressions l,m are coordinates on the sky, and u,v are the coordinates in spatial frequencies (baseline divided by wavelength). Detailed derivation of the above relation can be found in Clark (1989).

The spatial Fourier transform relation (1) holds for monochromatic radiation of a particular frequency. Therefore, in processing the measured fringe data, a temporal Fourier transform with respect to delay is often performed before the spatial Fourier inversion, making use of that fact that the fringe intensity in the delay direction satisfies a relation $I(t) = 2 I_0 [1 \pm V_1 \cos(2\pi v_0 t/\lambda + \varphi)]$ for uniform delay line scan velocity $v_0$, where $V_1$ is the normalized visibility amplitude and $\varphi$ the visibility phase. The wavelength-dependent phase shift allows the disentanglement of the various spectral components.

A two-element interferometer of baseline separation B in principle can measure spatial frequencies in the range (B-D, B+D), where D is the diameter of the single collector mirror. However, for a single-pointing observation, the spatial frequency obtained by a two-element interferometer is only that corresponding to baseline B. As shown in Ekers and Rots (1979), if one observes instead a series of pointings ($l_0$, $m_0$) and stamps the measured visibility by both the pointing and baseline coordinates, i.e. $V=V(u,v,l_0,m_0)$, then in principle one can obtain the flux in the entire range (B-D, B+D) through the use of the following Fourier transform relation with respect to the pointings ($l_0,m_0$):

$$\iint V(u,v,l_0,m_0) \cdot e^{-2\pi i(ul_0+vm_0)} dl_0 dm_0 \quad (2)$$
$$= [FT(A)(u,v)] \cdot [FT(B)(u_0-u, v_0-v)],$$

where $FT(A)(u,v)$, $FT(B)(u,v)$ are the Fourier transforms of $A(l,m)$ and $B(l,m)$, respectively, and where $V(u,v,l_0,m_0)$ is defined through the familiar relation between sky brightness and visibility, with the relevant quantities stamped by the pointing location of the primary beam

$$V(u,v,l_0,m_0) \equiv$$
$$\int_{-\infty}^{\infty}\int_{-\infty}^{\infty} A(l,m,l_0,m_0)B(l,m)e^{-2\pi i(ul+vm)} dldm. \quad (3)$$

In the measurement defined above, the visibility is measured with the interferometer having delay and phase reference center (0,0), and primary beam pointing center ($l_0$, $m_0$). Equation (2) can be used to evaluate FT(B) over an annulus around the baseline ($u_0,v_0$).

In practice, scanning for the continuous pointing range is not required, because the Fourier transform of the primary beam is band-limited. The Nyquist criterion tells us that $V(u,v,l_0,m_0)$ is fully specified by its values sampled on a regular grid in ($l_0,m_0$) with the grid spacing equal to $\lambda/2D$.

In the applications for SPECS, SPIRIT and WIIT, the multiple pointings are equivalently accomplished by a Nyquist-spaced large-format detector array on the focal plane, as long as delay is properly compensated for each pixel location.

The above scheme due to Ekers and Rots (1979) for a two element interferometer is often replaced in practice by the following alternative mosaic joint-deconvolution scheme for measurements using an interferometer array (Cornwell 1989). In this approach we maximize the entropy

$$H = -\sum_k B_k \ln \frac{B_k}{M_k \cdot e} \quad (4)$$

subject to the constraint of $\chi^2$:

$$\chi^2 = \sum_r \frac{|V(u_r,v_r,l_{0r},m_{0r}) - V'(u_r,v_r,l_{0r},m_{0r})|^2}{\sigma^2_{V(u_r,v_r,l_{0r},m_{0r})}} \quad (5)$$

where $B_k$ represents the model image brightness, $M_k$ represents the default image incorporating one's best prior knowledge of the recovered image, e is the base of the natural logarithm, $V(u_r, v_r, l_{0r}, m_{0r})$ is the observed visibility function at the baseline location ($u_r, v_r$) and sky location ($l_{0r}, m_{0r}$), $V'(u_r,v_r,l_{0r},m_{0r})$ is the model visibility distribution, and $\sigma_V$ is the rms noise of the measurement.

In this scheme, deconvolution is performed simultaneously with the untangling of information available at the various spatial frequencies. Since the required deconvolution is nonlinear, this approach yields better results especially for image points located near the overlapping boundaries of the different pointing positions. This approach also allows the addition of flux near zero spatial frequency measured in an autocorrelation mode. Mosaicing algorithms other than described above also exist in literature. We plan to evaluated the relative merit of each algorithm for dealing with a particular image reconstruction problem during the testbed operation.

## 5. CURRENT STATUS OF THE DESIGN AND CONSTRUCTION OF WIIT

WIIT has received full funding from NASA. We have by now completed the optical, mechanical, control and data system design, and this design has passed an internal peer review. The details of the testbed design is described in Zhang (2000a). Currently, the procurement process is essentially complete and the hardware components have mostly arrived. We are beginning the assembly process presently. In the mean time, data reduction and analysis software is being developed as a joint effort between Goddard and the University of Maryland (Zhang, Mundy and Teuben 2000).

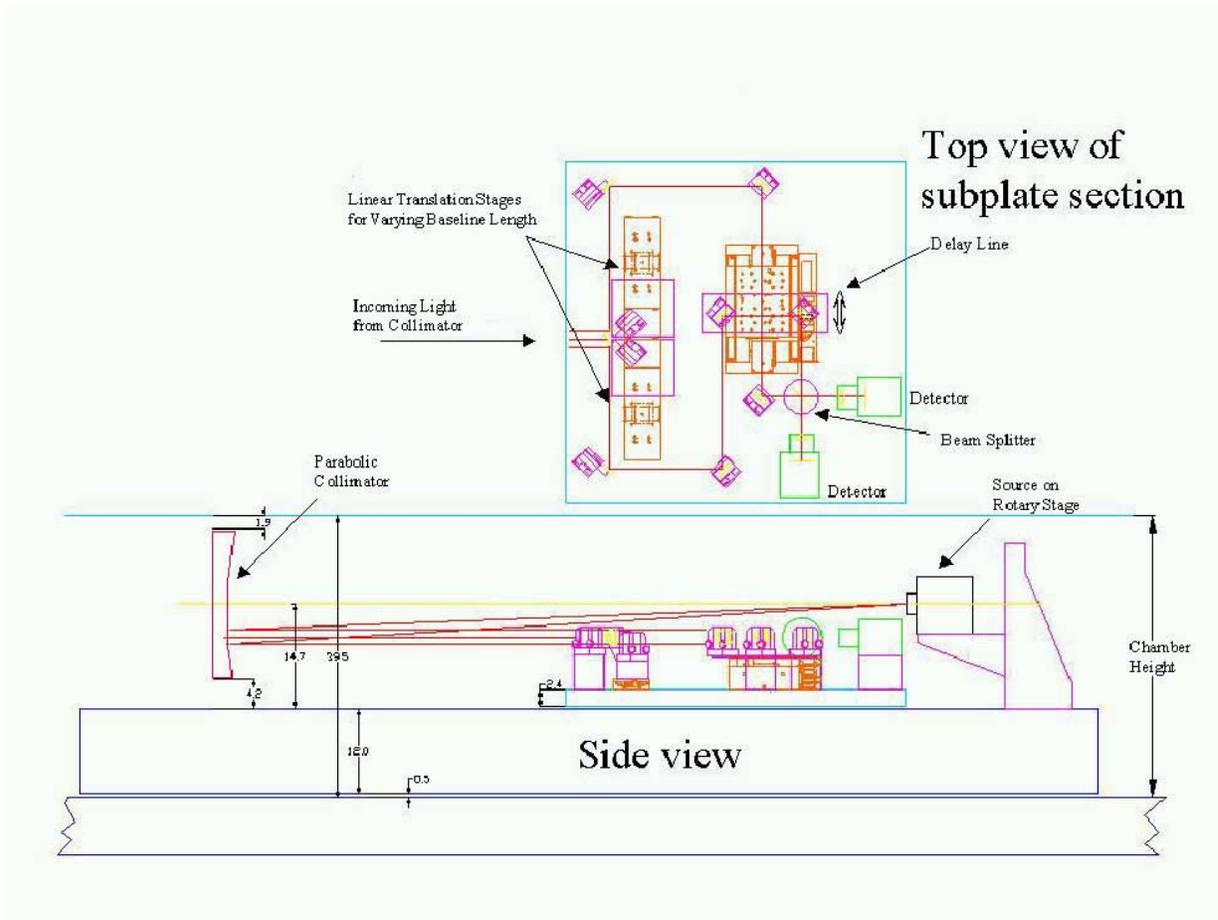

Figure 2: WIIT Layout

In Figure 2, we show the layout of the testbed. Since the ray-trace result (Zhang 2000b) showed that the curved design with beam reducers and Cat's eye delay lines did not introduce significant off-axis aberration, at least for the design parameters adopted for WIIT, we decided to start with an all-flat design as shown on this figure, which eliminated all curved mirrors (but kept the imaging lens). Investigations into how to deal with off-axis optics aberrations (which might well be present for the SPIRIT and SPECS design which involve much larger beam reduction ratios) will be left to the future when we upgrade the system. In the design that is chosen, we will have about 16 inches of usable baseline and 1 inch diameter collector mirrors. The focal length of the collimator paraboloid is about 2.4 meters. We will be using 100x100 - 400x400 detector pixels, and each pixel is further resolved by the interferometer into ~16x16 sub-pixels. The field of view ranges from 4 to16 arcminutes. We intend to achieve a spectral resolution on the order of $10^3$. The operating wavelength is about 0.6 μm. The mechanical tolerances of this optical wavelengths testbed exceed that of the future infrared interferometers. The experience gained here will help us understand the corresponding requirements for the space instruments.

In Figure 3, we show the schematics of the WIIT control and data systems. The testbed itself is controlled through a Labview interface by an industrial PC situated next to the optical bench. We will add remote-control and data monitoring capabilities as we pass the initial setup stage. A more sophisticated metrology system will be added if initial test results show a need for it. We expect to obtain the first fringes and show results of one-dimensional interferometry at the time of the conference.

## 6. ACKNOWLEDGMENTS

We thank the rest of the WIIT team at NASA's GSFC and at the University of Maryland for their contributions to the design and implementation of the testbed and the data analysis algorithms.

## 7. REFERENCES

Clark, B.G. 1989, "Coherence in Radio Astronomy", in Synthesis Imaging in Radio Astronomy, ASP Conf. Ser. vol. 6, eds. R.A. Perley, F.R. Schwab, & A.H. Bridle, `p.1`

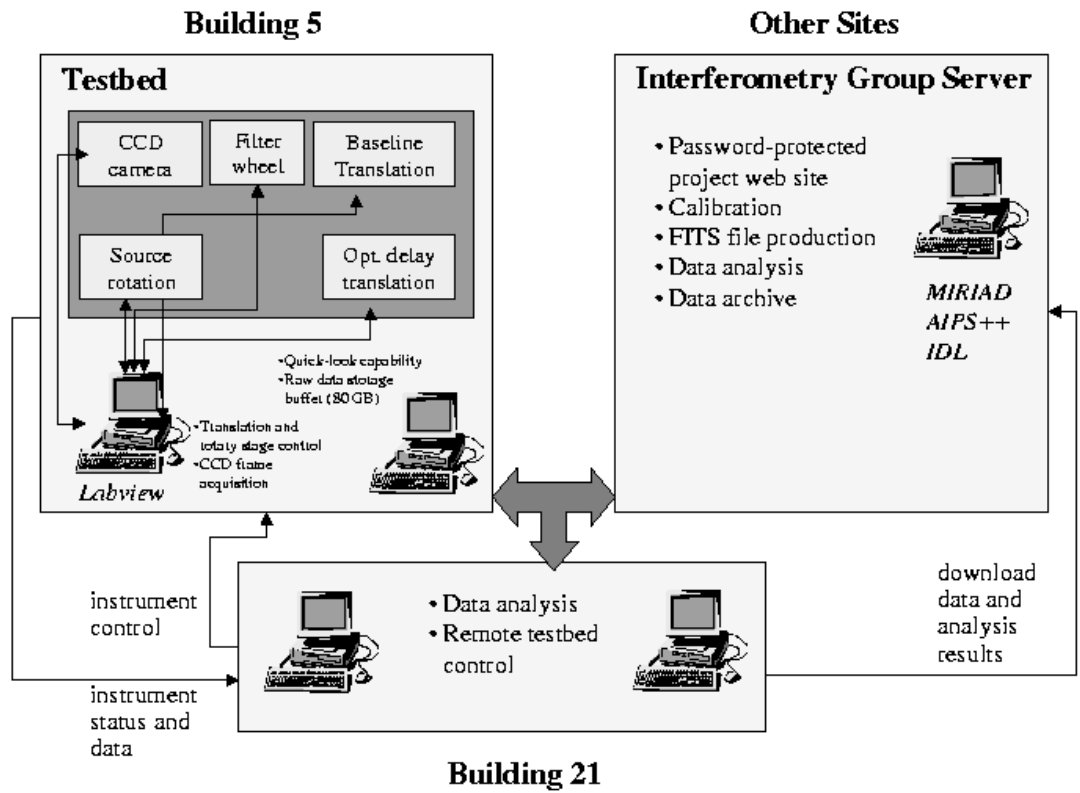

Figure 3: WIIT Control and Data System

Cornwell, T.J. 1989, "Wide Field Imaging III: Mosaicing", in *Synthesis Imaging in Radio Astronomy*, ASP Conf. Ser. vol. 6, eds. R.A. Perley, F.R. Schwab, & A.H. Bridle, p.277

Ekers, R.D., & Rots, A.H. 1979, "Short Spacing Synthesis from a Primary Beam Scanned Interferometer", Proc. IAU Coll. 49, *Image Formation from Coherence Functions in Astronomy*, C. van Schooneveld, Ed. (Dordrecht: Reidel), p. 61

Labeyrie, A. 1996, "Resolved Imaging of Extra-Solar Planets with Future 10-100 km optical Interferometer Arrays", Astro. Astrophys. Supp. Ser., 118, p. 517

Leisawitz, D. et al. 1999, "Wide-Field Imaging Interferometry", NASA proposal NRA-99-01-SPA-020.

Mather, J., Moseley, S. H., Leisawitz, Zhang, X., Dwek, E., Harwit, M., Mundy, L. G., Mushotzky, R. F., Neufeld, D., Spergel, D., & Wright, E. L. 2000, "The Submillimeter Probe of the evolution of cosmic structure", submitted to the *Review of Scientific Instruments*.

Traub, W. 1986, "combing Beams from Separated telescopes", Appl. Opt., Vol. 25, No. 4, p.528

Zhang, X. 2000a, "The Wide Field Imaging Interferometry Testbed: Objectives and Design", SPECS technical memo

Zhang, X. 2000b, "Diffraction and Aberration Characteristics of the WIIT Optics", SPECS technical memo

Zhang, X., Mundy, L., and Teuben, P. 2000, "The Application of Radio Astronomy Algorithms to Imaging Interferometry for SPECS SPIRIT and WIIT", in the *Proceedings of the Workshop on Computational Optics*, ed. R. Lyon

**Dr. Xiaolei Zhang** is Chief Scientist at Raytheon ITSS's Department for Astronomy and Solar Physics. Her research interests and experience include radio and far-infrared astronomical instrumentation, as well as the theoretical and observational studies of the formation and evolution of galaxies. She currently works at NASA's Goddard Space Flight Center on the design study of the space far-infrared interferometry missions.

**Lee Feinberg** was the Assistant Chief for Technology at the Instrument Technology Center of NASA's Goddard Space Flight Center. He had served as the co-lead engineering coordinator and technologist for the SPECS study team at GAFC. He had spent eight years in the instrument office for the Hubble Space Telescope Project, including serving as Instrument Manager for the Space Telescope Imaging Spectrograph and the Cosmic Origin Spectrograph. He has led the design of the Wide Field Camera 3 and co-developed the Detector Characterization Laboratory for WFC-3 at the GSFC. He was the technical lead responsible for interferometric verification of COSTAR and WFPC-II corrective aspheric optics.

**Dr. David Leisawitz** is Deputy Project Scientist for the COBE project at NASA's Goddard Space Flight Center. He is experienced in the fields of millimeter-wave molecular spectroscopy, radio astronomy, infrared photometry and IR data analysis. He conducts research on the interaction of massive starts with the interstellar medium. He leads the Far-IR Interferometry Mission Study Working Group and is the PI for WIIT.

**Douglas B. Leviton** is a senior optical scientist, who began his career at GSFC in 1983, has developed a wide variety of optical technologies and test methods for EUV through IR wavelengths for instruments for NASA science missions including the Cosmic Background Explorer, the Solar and Heliospheric Observatory, and the Hubble Space telescope. His patented ultra-high resolution, absolutee, optical encoder technology won the NASA Government Invention of the Year for 1999.

**Dr. Ahthony J. Martino** received his BS degree in physics from Xavier University in 1983, and his PhD in optics from the University of Rochester in 1990. Since 1990, he has been at NASA's Goddard Space Flight Center, where he is currently a group leader in the Lasers and Electro-Optics branch. He splits his time between electro-optical technology development projects and the design and construction of electro-optical instruments for space flight missions. Among the projects to which he has contributed are the Composite Infrared Spectrometer for Cassini, the COR1 coronagraph for STEREO, and the Geoscience Laser Altimeter System for ICESat.

**Dr. John C, Mather** is a senior astrophysicist at NASA's Goddard Space Flight Center. He has led the Cosmic Background Explorer satellite science effort from its proposal in 1974. As PI for the Far IR Absolution Spectrophotometer, he showed that the cosmic microwave background has a blackbody spectrum to 50 ppm. As the NASA Study Scientist for the Next Generation Space Telescope, he co-leads the science team with P. Stockman. He has many awards, and is a member of the National Academy of Sciences.

Notes added before submission to astro-ph: The assembly and alignment of WIIT optics was completed in Feburary 2001. First laser fringes was soon detected afterwards. A few more months was needed to track down and remove several sources of dispersion among the optical elements. On August 16, 2001, WIIT detected its first white light fringes, and is now proceeding to the imaging phase.